\documentclass[twocolumn,english,twocolumn,toolkits,showpacs]{revtex4}
\usepackage[T1]{fontenc}
\usepackage[latin9]{inputenc}
\usepackage{amsmath}
\usepackage{amssymb}
\usepackage{esint}

\makeatletter

\providecommand{\tabularnewline}{\\}

\@ifundefined{textcolor}{}
{%
 \definecolor{BLACK}{gray}{0}
 \definecolor{WHITE}{gray}{1}
 \definecolor{RED}{rgb}{1,0,0}
 \definecolor{GREEN}{rgb}{0,1,0}
 \definecolor{BLUE}{rgb}{0,0,1}
 \definecolor{CYAN}{cmyk}{1,0,0,0}
 \definecolor{MAGENTA}{cmyk}{0,1,0,0}
 \definecolor{YELLOW}{cmyk}{0,0,1,0}
 }

\usepackage{mathrsfs}\usepackage{amsfonts}\usepackage{bbm}\usepackage{graphics}\usepackage{epsfig}\@ifundefined{definecolor}
 {\@ifundefined{definecolor}
 {\usepackage{color}}{}
}{}
\usepackage{subfigure}\usepackage{chngcntr}\counterwithin{equation}{section}
\makeatother

\makeatother

\usepackage{babel}

\begin{document}

\title{On Chiral Symmetry Restoration at Finite Density in Large $\boldsymbol{N_{c}}$
QCD}

\date{$\today$}

\author{Prabal Adhikari}

\email{prabal@umd.edu}

\affiliation{Maryland Center for Fundamental Physics and the Department of Physics
University of Maryland, College Park, MD 20742}

\author{Raja R. M. Ayyagari}

\affiliation{Poolesville High School, Poolesville, MD 20837}

\author{Thomas D. Cohen}

\email{cohen@physics.umd.edu}

\affiliation{Maryland Center for Fundamental Physics and the Department of Physics
University of Maryland, College Park, MD 20742}

\author{Mark C. Strother}

\email{mcstro@umd.edu}

\affiliation{University of Maryland, College Park, MD 20742}

\begin{abstract}
At large $N_{c}$, cold nuclear matter is expected to form a crystal
and thus spontaneously break translational symmetry. The description
of chiral symmetry breaking and translational symmetry breaking can
become intertwined. Here, the focus is on aspects of chiral symmetry
breaking and its possible restoration that are by construction independent
of the nature of translational symmetry breaking---namely spatial
averages of chiral order parameters. A system will be considered to
be chirally restored provided all spatially-averaged chiral order
parameters are zero. A critical question is whether chiral restoration
in this sense is possible for phases in which chiral order parameters
are locally non-zero but whose spatial averages all vanish. We show
that this is not possible unless all chirally-invariant observables
are spatially uniform. This result is first derived for Skyrme-type
models, which are based on a non-linear sigma model and by construction
break chiral symmetry on a point-by-point basis. A no-go theorem for
chiral restoration (in the average sense) for all models of this type
is obtained by showing that in these models there exist chirally symmetric
order parameters which cannot be spatially uniform. Next we show that
the no-go theorem applies to large $N_{c}$ QCD in any phase which
has a non-zero but spatially varying chiral condensate. The theorem
is demonstrated by showing that in a putative chirally restored phase,
the field configuration can be reduced to that of a nonlinear sigma
model. It is also shown that this no-go theorem is fully consistent
with the vanishing of the spatial average of the chiral condensate
$\frac{1}{2}{\rm Tr}(U)$ (as happens in {}``half-skyrmion'' configurations).
This is because the chiral condensate is only one of an infinite set
of chiral order parameters, \textit{some} of which must be non-zero.
It is also shown that while an approximation of a unit cell of a Skyrme
crystal as a hypersphere does lead to a phase which is chirally restored
(in the average sense), this is an artifact of the approximation.
\end{abstract}
\maketitle

\section{Introduction}

One of the long-standing problems in QCD is to understand finite density
nuclear matter. Such matter is only poorly understood in terms of
QCD but is of fundamental importance in developing a theoretical understanding
of the qualitative and quantitative features of QCD equations of state
and phase diagrams. It is also important in studying astrophysical
objects such as neutron stars, which are composed of dense nuclear
matter, and in heavy ion collisions. While zero density nuclear matter
can be studied using lattice QCD, at finite densities or equivalently
chemical potential and low temperatures, the fermion sign problem
renders lattice studies intractable.

Given this situation one is often forced to rely on QCD-inspired models
to get insight. This is problematic. First of all, it is obvious that
the models are \textit{not} QCD and one needs to be cautious in interpreting
their results. Secondly, many models rely at least implicitly on QCD
being close to the large $N_{c}$ limit \cite{key-1,key-22} where
mean-field methods can be justified. At large $N_{c}$, quantum effects
become negligible and nuclear matter behaves classically in some important
sense. In this limit of high densities and large $N_{c}$, it is generally
believed that it is energetically favorable for nuclear matter to
crystallize in its ground state.

The large $N_{c}$ limit was recently used to motivate the existence
of a new phase of finite density nuclear matter. McLerran and Pisarski
\cite{key-3} argued that at low temperatures nuclear matter condenses
into a {}``quarkyonic phase'' of confined but chirally-restored
nuclear matter dominated by baryon-baryon interactions. This idea
was further developed in Ref. \cite{key-17}. This phase is supposed
to occur in large $N_{c}$ QCD in the regime in which the quark chemical
potential is of order $N_{c}^{0}$ large compared to $\Lambda_{{\rm QCD}}$.
The logic is simply that in this regime the quarks do not effect the
gluons and thus confinement described in terms of the Polyakov line
is unaffected. Thus, provided the chemical potential is large enough
to cause chiral restoration, one necessarily enters a quarkyonic phase.
This analysis is, strictly speaking, only clean in the chiral limit
of zero quark masses and for the purposes of this paper, we will ignore
the effects of quark masses and set them to zero at the outset.

There is a potential difficulty with the argument \cite{key-3}. \textit{A
priori} it is not trivially obvious from the structure of large $N_{c}$
QCD that at high baryon density, the system necessarily becomes chirally-restored.
This issue is at the crux of this paper. The primary question is whether
chiral symmetry is restored at high density. We have no answer to
this critical question. However, there is a secondary question of
importance which we can address, namely if chiral symmetry is restored,
by what mechanism does this occur?

To set the context for this, let us consider a model-dependent argument
used in Ref. \cite{key-3} to justify the assumption that chiral symmetry
is restored in large $N_{c}$ QCD at sufficiently high density. The
argument is based on the Skyrme model \cite{key-5,key-16} treated
at the classical level (which is justified at large $N_{c}$). The
model, although phenomenological, does capture many aspects of large
$N_{c}$ QCD and of chiral symmetry. It has proven to be a reasonable,
if crude way to describe static properties of baryons such as masses,
charge radii, and magnetic moments \cite{key-4}. It has been used
to describe nuclear matter which at mean-field level is naturally
described as a crystal \cite{key-6,key-7,key-8,key-10,key-14,key-15,key-20,key-12}.

At first sight, one might think that the Skryme model, which is based
on a non-linear sigma model and hence by construction breaks chiral
symmetry at every point in space, is unsuitable to study chiral restoration.
However, at this stage, it is worth noting that at low but non-zero
baryon density, two distinct types of symmetry are spontaneously broken:
chiral symmetry, which is broken as in the vacuum, and translational
symmetry, which is broken by the formation of a crystal. The two types
of spontaneous symmetry breaking can become intertwined. Since our
focus, is on chiral symmetry breaking and \textit{not} on translational
symmetry breaking it is reasonable to consider only those aspects
of chiral symmetry breaking which do not depend on translational symmetry
breaking.

It is easy to construct order parameters that are sensitive to chiral
symmetry breaking but insensitive to any details of translational
symmetry --- namely by spatially averaging standard chiral order parameters.
It is not unreasonable to consider a system chirally-restored if all
of the spatially averaged chiral order parameters vanish; this paper
will focus on the question of chiral restoration in this spatially
averaged sense. In the remainder of the paper, the phrase {}``chiral
restoration'' will often be used as a shorthand for {}``chiral restoration
in the spatial averaged sense''. One might hope that in Skyrme-type
models, chiral symmetry, although broken at every point in space,
could be restored in this spatially averaged sense.

In arguing that large $N_{c}$ QCD should restore chiral symmetry
at sufficiently high density, Ref. \cite{key-3} presents evidence
that the Skyrme model treated classically, in fact, exhibits chiral
restoration in this spatially-averaged sense and argues that large
$N_{c}$ QCD might similarly exhibit chiral restoration since the
Skyrme model captures many aspects of large $N_{c}$ QCD. The evidence
that the Skyrme model restores chiral symmetry at high density is
of two types: (i) the vanishing of the spatially-averaged chiral condensate
found in studies of different types of Skyrme crystals \cite{key-4,key-6,key-7,key-8,key-12,key-14,key-15}
and (ii) the vanishing of spatially-averaged chiral order parameters
as seen in analytical studies in which a Skymion in the crystal is
approximated as a single Skyrmion on the compactified manifold of
a hypersphere \cite{key-2,key-9,key-6,key-13}.

The evidence that chiral symmetry is restored in the spatially-averaged
sense at sufficiently high density in the Skyrme model appears to
be quite strong. However, as will be shown in this paper, it is misleading.
We prove a theorem that it is impossible for any Skyrme-type model
to restore chiral symmetry in the spatially-averaged sense at non-zero
baryon density. We further show that the no-go theorem also applies
to large $N_{c}$ QCD with a non-zero, spatially varying chiral condensate.
This is done by showing that in a chirally-restored phase, expectation
value of operators can be reduced to that of a generic nonlinear sigma
model.

Before proving this no-go theorem, it is useful to discuss how it
can possibly be correct given the evidence presented in Ref. \cite{key-3}.
First, let us consider the evidence of the vanishing of spatially-averaged
chiral condensate in Skyrme crystals above some critical density.
It is certainly possible that the chiral condensate itself may vanish
on spatial integration and indeed, it is expected to happen in so-called
\textquotedbl{}half-skyrmion\textquotedbl{} configurations of the
crystal's unit cell \cite{key-5,key-10,key-12,key-14,key-15}. This
occurs when isolated skyrmions (whose interaction energy at low density
scales as $r^{3}$) \cite{key-8}) are brought close together. It
is believed that lowest possible energy consistent with the topology
of the model (a winding number of unity per unit cell of the crystal)
is a new type of phase based on a configuration that has an additional
axial symmetry. This constrains the field to have $\sigma=-1$ at
the center of the unit cell. The configuration has a baryon number
of $\frac{1}{2}$ and can be transformed into another half-skyrmion
by the transformation $(\sigma,\mathbf{\boldsymbol{\pi})\rightarrow(-\sigma,-\mathbf{\boldsymbol{\pi})}}$.
The unit cell consists of two of these half-skyrmions. The chiral
order parameter $\frac{1}{2}{\rm Tr}(U)\equiv\sigma$, where $\sigma$
(which is assumed to be proportional to $\overline{q}q$ of QCD),
vanishes when integrated over the unit cell.

The key point is that while chiral restoration (in the average sense)
necessarily implies the vanishing of the spatially-averaged chiral
condensate, the converse is not true. It is logically possible that
chiral condensate could vanish due to the restoration of a discrete
symmetry while other spatially-averaged chiral order parameters remain
nonzero. \textit{A priori} it may seem implausible that such a scenario
could be realized. Indeed, it is reminiscent of Stern's suggestion
that in the vacuum, $\langle\overline{q}q\rangle$ vanishes but chiral
symmetry remains broken due to higher dimensional condensates\cite{key-21}.
However, as shown by Kogan, Kovner and Shifman (KKS) \cite{key-19}
using an elegant argument based on a rigorous QCD inequality for the
Euclidean space functional integral, the Stern scenario is inconsistent
with QCD. One might worry whether a similar scenario is also impossible
for finite density matter. However, the KKS argument depends on a
real and positive functional determinant, which while true for the
vacuum does not hold for finite density matter in which the chemical
potential leads to a functional determinant that is not manifestly
real and positive. Thus, this argument is not applicable to finite
density matter. Moreover, as will be shown in this paper, this unlikely
seeming scenario must occur for both Skyrme models and for large $N_{c}$
QCD if there exists a regime where the average chiral condensate is
not zero locally but vanishes under spatial averaging.

Now let us consider the second class of evidence for chiral restoration
in the average sense---namely the fact that when a Skyrmion in the
crystal is approximated as a single Skyrmion on the compactified manifold
of a hypersphere, all spatially-averaged chiral order parameters vanish.
In effect by using Skyrmions placed on the compactified manifold of
a hypersphere {}``surface'', $S^{3}(L)$ embedded in $R^{4}$ \cite{key-2,key-9,key-6,key-11},
one is effectively chopping off the {}``edges'' of a unit cell in
a real Skyrme crystal and somehow mapping this onto the hypersphere
again with the winding number set to one. Of course, in doing this
one is making an uncontrolled approximation, but the virtue is simplicity
with many quantities analytically calculable. The density of the Skyrme
crystal is assumed to be the inverse volume of the hypersphere. For
low densities (or large radius hypersphere), there are two solutions---a
stable low energy one in which the winding number density (\textit{i.e.,}
baryon density) is localized and spatially-averaged chiral order parameters
are nonzero and an energetically unstable phase in which the winding
number density is spread uniformly and spatial averages of all chiral
order parameters vanish. However at some critical density (or equivalently
radius) the two solutions merge and above this density the lowest
energy solution is the one with uniform baryon density and chiral
restoration in the spatially-averaged sense.

Of course, this evidence for chiral restoration depends on an uncontrolled
approximation. One does not know at the outset how well one can approximate
a Skyrmion in the crystal by a single Skyrmion on a hypersphere. Presumably,
the logic underlying the approximation is that the principal effect
of putting a Skyrmion into a crystal is to restrict the space over
which it can spread. If this is the case it is natural to assume that
a Skyrmion in the restricted space of a hypersphere is likely to be
qualitatively similar to a Skyrmion in a crystal. Thus, it is plausible
that while details of such a precise equation of state or the density
of the phase transition to the chirally-restored phase would be affected
by the approximation, qualitative features such as the existence of
a chirally-restored phase would not.

The preceding argument depends on a critical assumption, namely that
using a hypersphere to restrict the volume of the Skyrmion acts generically
like other restrictions on its volume. This need not be true. The
hypersphere is a rather special geometrical structure. One might worry
that the chiral restoration seen on the hypersphere is not a generic
feature at all, but rather an accidental feature of the \textit{ad
hoc} choice of geometry. If this is the case then the fact that Skyrmions
exhibit chiral restoration at high density on the hypersphere gives
no insight into what happens for Skyrme crystals. As will be shown,
the vanishing of all spatially-averaged chiral order parameters \textit{is}
a result of the peculiarities of the geometry of the hypersphere.
This can be seen clearly by keeping the Skyrmion on a compact space
but one in which the geometry is distorted away from the hypersphere.
When this is done, chiral restoration (in the average sense) disappears.

The paper is organized as follows. In the following section, we prove
a no-go theorem that prevents chiral symmetry restoration (in the
average sense) for all Skyrme-type models, \textit{i.e.}, non-linear
sigma models in which the winding number density is identified as
the baryon density. This proof consists of two parts. First, we derive
as a necessary condition for chiral restoration (in the average sense)
the condition that all chiral singlet observables must be spatially
uniform. Next, we show that a particular chirally symmetric order
parameter cannot be uniform while chiral symmetry is restored simultaneously
in Skyrme-type models, thus establishing the theorem. Then we extend
our no-go theorem from the realm of models to large $N_{c}$ QCD itself.
In particular, we prove that large $N_{c}$ QCD at non-zero baryon
density cannot have a chiral condensate, which is non-zero but non-uniform
spatially in such a way that all spatially-averaged chiral order parameters
vanish. In the following section, we discuss in detail the case of
a Skyrmion confined to a hypersphere in light of the no-go theorem.
We show explicitly why the proof of the no-go theorem works for crystals
in flat space but breaks down for a Skyrmion confined to a hypersphere.
Next, we discuss why the insights gleaned from a Skyrmion confined
from a hypersphere is misleading. We do this by considering Skyrmions
confined to a compact manifold which is deformed from a hypersphere;
chiral symmetry in the average sense is not restored. This demonstrates
that chiral restoration observed on the hypersphere is an artifact
of the (unphysical) choice of geometry. We end with a brief discussion
of these results.

\section{A No-go Theorem in the Skyrme Model\label{2}}

In this section, we prove a no-go theorem for Skyrme-type models,
which we define as chirally invariant non-linear sigma models based
on a matrix-valued field: $U(\mathbf{x})$ with $U\in SU(2)$ (and
possibly other fields) with a Lagrangian rich enough to support stable
topological solitons. The model is treated classically, in keeping
with large $N_{c}$ QCD. The classical field, $U(\mathbf{x})$, which
we take to be continuous is a mapping from $R^{3}\rightarrow SU(2)$
for flat space (and from $S^{3}(L)\rightarrow SU(2)$ for skyrmions
on the hypersphere). Physically, the field $U$ encodes the dynamics
of the pion. Such models automatically have an algebraically conserved
winding number current, $w^{\mu}$ (that is $\partial_{\mu}w^{\mu}=0$
for any field configuration) where \begin{equation}
w^{\mu}=\frac{\epsilon^{\mu\nu\alpha\beta}\,{\rm Tr}\left[(U^{\dagger}\partial_{\nu}U)(U^{\dagger}\partial_{\alpha}U)(U^{\dagger}\partial_{\beta}U)\right]}{24\pi^{2}}\label{winding}\end{equation}
In these models, $w^{\mu}$ is often taken to be to equal to the conserved
baryon current, although in principle they can differ by a quantity,
which is conserved and has zero net charge.

In proving the no-go theorem, the precise details of the Lagrangian
play no role nor does the identification of the winding number current
with baryon current. However, its chiral properties are essential.
The Lagrangian must be invariant under a global $SU(2)_{L}\times SU(2)_{R}$
transformation where $U(\mathbf{x})\rightarrow LU(\mathbf{x})R^{\dagger}$
and $U^{\dagger}(\mathbf{x})\rightarrow RU^{\dagger}(\mathbf{x})L^{\dagger}$.
If the theory has other fields, they must transform in a consistent
way. Of course, the non-linear sigma models build in spontaneous symmetry
breaking at the outset; while the Lagrangian is chirally invariant,
any field configuration at any point in space is not. That is $U(\mathbf{x})\neq RU(\mathbf{x})L^{\dagger}$.
Thus, by construction, chiral symmetry when evaluated point-by-point
in space \textit{cannot be restored} in any model in this class.

The notion of chiral symmetry restoration in the Skyrme model only
makes sense as a global property and not a local one. We define chiral
symmetry to be restored in the spatially-averaged sense if \textit{all}
functions of the classical fields when spatially averaged over a unit
cell equals the average of the same function averaged uniformly over
the internal space of chiral rotations. One particularly interesting
class of observables are those that depend on $U$ (which already
encodes spontaneous symmetry breaking) but not on spatial derivatives
(or other possible fields in the problem) where one can project out
the chiral singlet part by averaging uniformly over the $SU(2)$ field.
Thus, for example, chiral restoration in this sense requires that
any scalar function of $U(\mathbf{x})$ only (and not its derivatives)
will satisfy 
\begin{equation}
\frac{1}{V_{R^{3}}}\int_{R^{3}}dV\, F(U(\mathbf{x}))=\frac{1}{2\pi^{2}}\int_{SU(2)}d\mu\, F(U(\mu))\;,\label{eq:2}
\end{equation}
where $F$ is an arbitrary real-valued function of $U$ and $\mu$
represents the three angles needed to specify a general $SU(2)$ matrix
and $d\mu$ is the Haar measure of $SU(2)$. The logic of this is
quite simple: averaging over the internal space ensures that all chiral
order parameters will vanish while chiral singlets will not. Note
that in Eq.~(\ref{eq:2}) the average of the function of the field
$U(\mathbf{x})$ in the internal space is independent of the dynamics
of the model, but its spatial average depends on the equations of
motion.

In this section, we prove that chiral restoration in the spatially-averaged
sense is impossible for the Skyrme model. This proof has two parts.
The first is a demonstration that chiral restoration in this sense
cannot occur unless all chiral singlet observables are spatially uniform.
The second is that chiral restoration is inconsistent with the fact
that all chiral order parameters must vanish (upon spatially averaging).
This will be demonstrated via the explicit construction of a chiral
singlet operator which cannot be spatially uniform in a regime where
the spatial average of all chiral order parameters vanish.

\subsection{Chiral restoration in the average sense requires spatial uniformity
for all chiral singlets \label{2a}}

The key to the first part of the proof is the intuition that the only
natural way chiral restoration in the average sense can occur is if
spatially integrating corresponds to integrating over the internal
space with uniform weighting. The central part of the proof is a demonstration
that this intuition is correct. Ultimately, this proves to be such
a strong constraint that no field distributions in Skyrme-type models
can satisfy it unless all chiral singlet observables are spatially
uniform.

To begin the formal treatment of the first part of the proof, we note
that chiral restoration requires \textit{all} chiral order parameters
vanish, or equivalently that the spatial average of \textit{all} functions
equals the uniform average over all chiral rotations. Thus, to show
that chiral restoration cannot occur it is sufficient to show that
there exists some subset of chiral order parameters for which it is
not true. Here, we will focus on observables constructed entirely
from the field $U$, which is local and hence depends on a single
point $\mathbf{x}$, as in Eq.~(\ref{eq:2}). Examples of this class
of observables include $F(U)={\rm Tr}[U]$, and $F(U)={\rm Tr}[U\tau_{1}]^{2}+{\rm Tr}[U]^{2}$.
Note that function $U$ is completely specified by three Euler angles,
denoted collectively as $\mu$ so that the most general $F$ is simply
the most general function $\mu$ consistent with periodicity conditions.
The Wigner D-matrices provide a complete basis of functions, which
satisfy these boundary conditions so that we can always decompose
$F\left(U(\mu)\right)$ into the following form:
\begin{equation}
F\left(U(\mu)\right)=\sum_{j,k}f_{m,k}^{j}D_{m,k}^{j}(\mu)\,,\label{decomp}
\end{equation}
where the coefficients $f_{m,k}^{j}$ are constants which depend only
on the function $F$. Note that the mapping from physical space onto
$U$ can be recast as a mapping from real space onto $\mu$: $F\left(U(\mathbf{x})\right)=F\left(U(\mu(\mathbf{x}))\right)$
We will show that for any smooth field configuration, there must be
members of this class of observable which do not satisfy Eq.~(\ref{eq:2})
unless all chiral singlet observables are spatially uniform.

To proceed, it is useful to recall that the Wigner D-matrices satisfy
an orthogonality condition
\begin{equation}
\int d\mu D_{m',k'}^{j'}D_{m,k}^{*j}=\frac{8\pi^{2}}{2j+1}\delta_{m'm}\delta_{j'j}\delta_{k'k}\,.\label{eq:2-1}
\end{equation}
 Since, the D-matrix for $j=0$ is a constant independent of $\mu$,
it follows from the orthogonality relation that for $j\neq0$,
\begin{equation}
\int d\mu D_{m,k}^{j}=0\,.
\end{equation}
Thus, it is clear that $F$ is a chiral order parameter (\textit{i.e.,}
an observable which is necessarily zero if chiral symmetry is unbroken)
if, and only if, when written in the form of the decomposition of
Eq.~(\ref{decomp}) the coefficient $f_{0,0}^{0}=0$.

Since the Wigner-D matrices form a basis, Eq.~(\ref{eq:2}) can only
be satisfied for \textit{all} choices of $F$ if it is satisfied for
each Wigner-D matrix. A simple way to implement this is to simply
choose $F=D_{m,k}^{j}\,(j\neq0)$. Equation~(\ref{eq:2-1}) then
implies that if chiral restoration in the average sense occurs, then
\begin{equation}
\int F\left(U(\mu(\mathbf{x}))\right)dV=\int D_{m,k}^{j}J(\mu)d\mu=0\,,\label{jac}
\end{equation}
 where $J(\mu)$ is the determinant of the Jacobian for a transformation
from the physical space onto the internal space and the right-hand
side of Eq. (\ref{eq:2-1}) is zero by orthogonality provided that
$j\ne0$. Note that the mapping of $\mathbf{x}$ onto $\mu$ need
not be one-to-one but one can always cast the integral into the preceding
form by combining the Jacobian determinants for any regions in real
space which are mapped into the same region of internal space.

The next step is simply to note that since the Jacobian determinant
is a scalar function of $\mu$, it too can be decomposed into a basis
of Wigner-D matrices 
\begin{equation}
J(\mu)=\sum_{j',m',k'}c_{m',k'}^{j'}D_{m',k'}^{*j'}(\mu).
\end{equation}
Here, we have used complex conjugates of the Wigner matrices to exploit
the orthogonality condition of Eq. (\ref{eq:2-1}). Note that the
coefficients $c_{m',k'}^{j'}$ characterize the field configuration
and are independent of the observable $F$. Inserting this decomposition
into Eq.~(\ref{jac}) for $F=D_{m,k}^{j}$ and using orthogonality
implies that if chiral restoration in the average sense occurs only
if $c_{m,k}^{j}=0$ for all $j\ne0$.

Note that the choice of $F$ is arbitrary and thus we we can repeat
the analysis for all $j$, $m$ and $k$ and thus deduce that if chiral
symmetry is restored then \textit{all} $c_{m,k}^{j}=0$ except when
$j=0$ (which corresponds to a Jacobian independent of $\mu$). Hence
the determinant of the Jacobian is a constant. This in turn implies
that chiral restoration in the average sense means that spatially
averaging must be equivalent to a uniform averaging over the internal
space, precisely as one would expect intuitively.

To proceed further, let us consider a broader class of order parameter:
$G\left(U(\mu(\mathbf{x}))\right)S(\mathbf{x})$, where $G$ is a
chiral order parameter of the class considered previously, namely
a function of $U$ only and $S$ is a chiral scalar of $U$ (but may
be non-local e.g., $S(\mathbf{x})=U^{\dagger}(\mathbf{x})U(\mathbf{x}+\mathbf{d})$,
with $\mathbf{d}$ a vector-valued fixed point, or may potentially
also depend on derivatives or integrals e.g.,~$\int d^{3}y\,{\rm Tr}[U^{\dagger}(\mathbf{x})U(\mbox{\textbf{x}+\textbf{y}})]\exp(-\frac{y^{2}}{L^{2}})$.
It is easy to see that a necessary condition for chiral restoration
in the average sense is 
\begin{equation}
\begin{split} & \frac{1}{V_{R^{3}}}\int_{R^{3}}G(\mu(\mathbf{x}))S(\mathbf{x})\, dV=\\
 & \left(\frac{1}{2\pi^{2}}\int_{SU(2)}G(\mu)\, d\mu\right)\left(\frac{1}{V_{R^{3}}}\int_{R^{3}}S(\mathbf{x})\, dV\right)\,.\end{split}
\label{eq:1}
\end{equation}
This follows since if the system is chirally restored in the average
sense, chiral rotations do not affect the integral on the left-hand
side. Thus, we can chirally average without affecting the result.
On the other hand, $S(\mathbf{x})$ is unaffected by chirally averaging.
Moreover, when averaging over chiral space for $G$, it does not matter
where in chiral space one starts, one gets the same result; thus the
chiral average is independent of the value of $U(\mathbf{x})$ and
hence of \textbf{$\mathbf{x}$}. One can therefore remove the chirally-averaged
$G$ from the spatial average yielding Eq.~(\ref{eq:1}).

Since $G$ is an order parameter, the first factor on the right-hand
side is necessarily zero and the integral on the left-hand side must
vanish. Thus 
\begin{equation}
\begin{split} & \frac{1}{V_{R^{3}}}\int_{R^{3}}G\left(U(\mu(\mathbf{x}))\right)S(\mathbf{x})\, dV=\\
 & \int G(\mu)S(\mathbf{x}(\mu))\, J(\mu)d\mu=0\,,\end{split}
\end{equation}
where the second form follows from change variables of integration
to $\mu$; as before we can always cast the integral into this form
even if the mapping is not one-to-one with one region in $\mu$ corresponding
to more than one region in $\mathbf{x}$ by summing over the contributions
of the Jacobian time $S$ for the different regions in $\mathbf{x}$.
At this stage we can simply repeat the argument used previously for
$F$ to show that chiral restoration in the average sense requires
that $S(\mathbf{x}(\mu))J(\mu)$ must be a constant independent of
$\mu$. On the other hand, we have already shown that $J$ is a constant.
Thus, we conclude that $S$ also must be a constant.

Note that this is a very strong constraint. Since chiral restoration
means that \textit{all} chiral order parameters must vanish and since
the previous argument holds for any chiral scalar operator, we conclude
that in order for chiral restoration in the average sense to occur,
\textit{all chiral scalars} must be uniform in space. This completes
the first part of the proof.

\subsection{Spatial uniformity for all chiral singlets is incompatible with chiral
restoration}

The next step is to show that in Skyrme-type models it is not possible
for all chiral singlets to be spatially uniform while simultaneously
having chiral symmetry restored (in the average sense). We do so via
an indirect proof: we will assume that the system is in a chirally
restored phase and then show that if all chiral singlets are spatially
uniform, there is a mathematical inconsistency.

A key point about chirally restored phases that was stated in the
previous section is that spatially averaging is equivalent to averaging
uniformly over the internal space. This in turn means that every value
of $U$ occurs at some point in space (and indeed in a crystal at
least once per unit cell). Thus, there must exist four points $\mathbf{x}_{0}$,
$\mathbf{x}_{1}$, $\mathbf{x}_{2}$ and $\mathbf{x}_{3}$ with the
property that 
\begin{equation}
\begin{split}U(\mathbf{x}_{0}) & =I\\
U(\mathbf{x}_{1}) & =i\tau_{1}\\
U(\mathbf{x}_{2}) & =i\tau_{2}\\
U(\mathbf{x}_{3}) & =i\tau_{3}\end{split}
\end{equation}
 where $I$ is the two-dimensional identity matrix.

To proceed further, we introduce three matrix-valued functions: 
\begin{equation}
\begin{split}\theta_{1}(\mathbf{x}) & =U(\mathbf{x})U^{\dagger}(\mathbf{x}+\mathbf{x}_{1}-\mathbf{x}_{0})\\
\theta_{2}(\mathbf{x}) & =U(\mathbf{x})U^{\dagger}(\mathbf{x}+\mathbf{x}_{2}-\mathbf{x}_{0})\\
\theta_{3}(\mathbf{x}) & =U(\mathbf{x})U^{\dagger}(\mathbf{x}+\mathbf{x}_{3}-\mathbf{x}_{0})\end{split}
\end{equation}
Next observe that under chiral transformations $\theta_{j}(\mathbf{x})$
transforms according to $\theta_{j}(\mathbf{x})\rightarrow L\theta_{j}(\mathbf{x})L^{\dagger}$
where $L$ is the SU(2) matrix generating left chiral rotations. Note
that by construction \begin{equation}
\theta_{j}(\mathbf{x}_{0})=-i\tau_{j}\,.\label{tau}\end{equation}
 The analysis in this section is based on the chiral properties of
quantities constructed from $\theta_{j}$.

It is obvious from the transformation properties that ${\rm Tr}\left(\theta_{j}(\mathbf{x})\right)$
is a chiral invariant. Since we are in a chirally restored phase in
the average sense (by hypothesis), this requires ${\rm Tr}\left(\theta_{j}(\mathbf{x})\right)$
to be translationally invariant. However, by construction ${\rm Tr}\left(\theta_{j}(\mathbf{x}_{0})\right)=0$,
from which one sees that ${\rm Tr}\left(\theta_{j}(\mathbf{x})\right)=0$
for all $\mathbf{x}$. Since the $\theta_{j}(\mathbf{x})$ are generically
traceless SU(2) matrices, they can be written in the following form:
\begin{equation}
\theta_{j}(\mathbf{x})=-i\sum_{a}n_{j}^{a}(\mathbf{x})\tau_{a}\;\;{\rm where}\;\; n_{j}^{a}(\mathbf{x})=i\frac{{\rm Tr}\left(\theta_{j}(\mathbf{x})\tau^{a}\right)}{2}\;.
\end{equation}
 For our purposes, it is useful to consider the $n_{j}^{a}$ as being
components of three distinct unit isovectors each labeled by $j$.
Thus $\hat{n}_{j}(\mathbf{x})\equiv(n_{j}^{1}(\mathbf{x}),n_{j}^{2}(\mathbf{x}),n_{j}^{3}(\mathbf{x}))$
so that $\theta_{j}(\mathbf{x})=-i\hat{n}_{j}(\mathbf{x})\cdot\vec{\tau}$
where arrows will indicate vectors in isospace and hats unit vectors
in isospace. The fact that $\hat{n}_{j}(\mathbf{x})$ has unit norm
follows from the fact that $\theta_{j}(\mathbf{x})\in{\rm SU(2)}$.

A very useful set of chirally invariant quantities can be constructed
from the $\hat{n}_{j}(\mathbf{x})$: 
\begin{equation}
\begin{split}W_{ij}^{\mathbf{d}}(\mathbf{x}) & \equiv\hat{n}_{i}(\mathbf{x})\cdot\hat{n}_{j}(\mathbf{x}+\mathbf{d})\\
 & =-\frac{{\rm Tr}\left(\theta_{i}(\mathbf{x})\,\theta_{j}(\mathbf{x}+\mathbf{d})\right)}{2}\,,\end{split}
\label{W}
\end{equation}
where $\mathbf{d}$ is a vector-valued parameter. It is quite straightforward
to show that for any values of $i$,$j$ and $\mathbf{d}$, $W_{ij}^{\mathbf{d}}$
is, in fact, a chiral singlet. Note that given the result demonstrated
in the previous section, \textit{if} the system is in a chirally-restored
phase, then $W_{ij}^{\mathbf{d}}(\mathbf{x})$ must be independent
of $\mathbf{x}$ for fixed $i$,$j$ and $\mathbf{d}$.

Clearly, by choosing different values of $\mathbf{d}$, one can probe
spatial correlations among the $L_{i}$. One important case is where
$\mathbf{d}=\mathbf{0}$. For notational convenience we will introduce
a special symbol for this: \begin{equation}
Q_{ij}(\mathbf{x})\equiv W_{ij}^{\mathbf{0}}(\mathbf{x})=\frac{{\rm Tr}\left(\theta_{i}(\mathbf{x})\,\theta_{j}(\mathbf{x}+\mathbf{d})\right)}{2}\;.\label{Q}\end{equation}
 It is trivial to see from Eq.~(\ref{Q}) and Eq.~(\ref{tau}) that
\begin{equation}
Q_{ij}(\mathbf{x}_{0})=\delta_{ij}\,.
\end{equation}
By hypothesis, the system is in a chirally-restored phase and thus
$Q_{ij}(\mathbf{x})$ is a chiral singlet independent of $\mathbf{x}$.
Since $Q_{ij}$ is simply the inner product of a set of three unit
isovectors, the most general way for it to be independent of position
is if the set of $\hat{n}$ at one point are related to the set at
another by the most general inner product conserving transformation,
an element of SO(3). Thus we conclude that 
\begin{equation}
\hat{n}_{j}(\mathbf{x})=\overleftrightarrow{R}(\mathbf{x})\hat{n}_{j}(\mathbf{x}_{0})\;\;{\rm with}\overleftrightarrow{R}(\mathbf{x}_{0})\in{\rm SO}(3)\;,\label{R_0}
\end{equation}
where $\overleftrightarrow{R}(\mathbf{x})$ is the same for all $j$.
Note that assuming $U$ describes a configuration, which is chirally
restored in the average sense, $\overleftrightarrow{R}(\mathbf{x})$
is completely determined by the chiral field $U(\mathbf{x})$. Recall
that $\theta_{j}(\mathbf{x})$ transforms according to $\theta_{j}(\mathbf{x})\rightarrow L\theta_{j}(\mathbf{x})L^{\dagger}$
and are thus in the irreducible ($\frac{1}{2}$,$\frac{1}{2}$) chiral
representation. As such they are chiral order parameters and by definition
their spatial averages must vanish in a chirally-restored regime.
This in turn means the spatial average of the $\hat{n}_{j}$ must
be zero. Using Eq.~(\ref{eq:2}) this means that in a chirally restored
phase 
\begin{equation}
\frac{1}{V_{R^{3}}}\int_{R^{3}}dV\overleftrightarrow{R}(\mathbf{x})\hat{n}_{j}(\mathbf{x}_{0})=0
\end{equation}
 for all $j$. However since the $\hat{n}_{j}$ form a complete and
orthonormal basis, this is possible only if the integral of the matrix
itself vanishes: 
\begin{equation}
\frac{1}{V_{R^{3}}}\int_{R^{3}}dV\overleftrightarrow{R}(\mathbf{x})=0\label{Rav}
\end{equation}

The hypothesis that the system is in a chirally-restored phase (in
the spatially-averaged sense) requires that all chiral singlet observables
are independent of position. Since $W_{ij}^{\mathbf{d}}(\mathbf{x})$
is chiral singlet, it must be independent of $\mathbf{x}$. Using
Eq.~(\ref{R_0}) and the definition of $W$ from Eq.~(\ref{W})
we see that 
\begin{equation}
W_{ij}^{\mathbf{d}}(\mathbf{x})=\hat{n}_{i}^{T}(\mathbf{x}_{0})\overleftrightarrow{R}^{T}(\mathbf{x})\overleftrightarrow{R}(\mathbf{x+d})\hat{n}_{j}(\mathbf{x}_{0})\label{WR}
\end{equation}
 Note that $W_{ij}^{\mathbf{d}}(\mathbf{x})$ can be expressed as
the matrix $\overleftrightarrow{R}^{T}(\mathbf{x})\overleftrightarrow{R}(\mathbf{x}+\mathbf{d})$
evaluated between the vectors $\hat{n}_{j}(\mathbf{x}_{0})$. However,
since these vectors form a complete basis and are independent of $\mathbf{x}$,
the only way for $W_{ij}^{\mathbf{d}}(\mathbf{x})$ to be independent
of $\mathbf{x}$ is if the matrix $\overleftrightarrow{R}^{T}(\mathbf{x})\overleftrightarrow{R}(\mathbf{x+d})$
itself is also independent of $\mathbf{x}$: 
\begin{equation}
\overleftrightarrow{R}^{T}(\mathbf{x})\overleftrightarrow{R}(\mathbf{x}+\mathbf{d})=\overleftrightarrow{R}(\mathbf{d}+\mathbf{x}_{0})\;.
\end{equation}
 But since the transpose of a rotation matrix is its inverse, 
\begin{equation}
\overleftrightarrow{R}(\mathbf{x}+\mathbf{d})=\overleftrightarrow{R}(\mathbf{x})\overleftrightarrow{R}(\mathbf{d}+\mathbf{x}_{0})\;.\label{RR}
\end{equation}
 Note that $\mathbf{x}$ and $\mathbf{d}$ are arbitrary three-dimensional
vectors in space. Thus the form of Eq.~(\ref{RR}) holds under replacements
of $\mathbf{x}$ and $\mathbf{d}$ by other vectors. Consider in particular
the replacements $\mathbf{x}\rightarrow\mathbf{d}+\mathbf{x}_{0}$
and $\mathbf{d}\rightarrow\mathbf{x}-\mathbf{x}_{0}$. Thus $\overleftrightarrow{R}(\mathbf{x}+\mathbf{d})=\overleftrightarrow{R}(\mathbf{d}+\mathbf{x}_{0})\overleftrightarrow{R}(\mathbf{x})$.
Equating the two ways of writing $\overleftrightarrow{R}(\mathbf{x}+\mathbf{d})$
and relabeling $\mathbf{d}+\mathbf{x}_{0}$ as $\mathbf{y}$ yields
\begin{equation}
[\overleftrightarrow{R}(\mathbf{y}),\overleftrightarrow{R}(\mathbf{x})]=0\label{com}
\end{equation}
 Note that since $\mathbf{d}$ was arbitrary so is $\mathbf{y}$ and
Eq.~(\ref{com}) holds for all values of $\mathbf{x}$ and $\mathbf{y}$.
Eq.~(\ref{com}) strongly constrains the nature of $\overleftrightarrow{R}(\mathbf{x})$:
$\overleftrightarrow{R}$ at any points commutes with $\overleftrightarrow{R}$
at any other point while general three-dimensional rotations do not.
Indeed, the only way that Eq.~(\ref{com}) can be satisfied for all
points is if the rotations at every point in space are all about a
common spatially independent rotation axis in isospace. Let us denote
the unit vector in the direction of this axis $\hat{n}_{{\rm rot}}$.
It is a general property of rotation matrices in three dimensions
that the axis of rotation is an eigenvector with an eigenvalue of
unity. Since $\hat{n}_{{\rm rot}}$ is independent of $\mathbf{x}$
it is also an eigenvector of the spatially-averaged rotation matrix
again with an eignenvalue of unity: 
\begin{equation}
\frac{1}{V_{R^{3}}}\int_{R^{3}}dV\overleftrightarrow{R}(\mathbf{x})\hat{n}_{{\rm rot}}=\hat{n}_{{\rm rot}}\;.\label{eigen}
\end{equation}

However, Eq.~(\ref{eigen}) is clearly inconsistent with Eq.~(\ref{Rav}):
the spatially-averaged matrix cannot simultaneously be zero and have
an eigenvalue of one. Note that both Eq.~(\ref{eigen}) and Eq.~(\ref{Rav})
are direct mathematical consequences of our assumption that the system
is in a chirally-restored phase (in the spatially-averaged sense).
Equation (\ref{Rav}) followed quite directly from this as by definition
all chiral order parameters, including $\theta_{j}$, must vanish
upon spatially averaging in this phase. Equation (\ref{eigen}) followed
from the requirement proved in the previous subsection that all chiral
singlets, including $W_{ij}^{\mathbf{d}}$, must be spatially uniform
in such a phase. Since these two are inconsistent, we have shown that
the assumption underlying them---that the system is in a phase which
is chirally restored in a spatially averaged sense---cannot be correct.
Moreover, since the analysis was done for an arbitrary field configuration
for the entire class of Skyrme-type models we have proved that chiral
restoration in the average sense cannot occur in models of this class. 

Ultimately, this result should not be surprising. The only natural
way to conceive of a situation in which chiral order parameters to
be non-zero locally but zero under spatial averaging is for spatial
averaging to uniformly cover all directions in the internal chiral
space. This will occur if the field configurations map from real space
to chiral space in such a way that uniform coverage in one is mapped
to uniform coverage in the other. However, the internal chiral space
is curved while the physical space is flat and hence such a mapping
is not possible.

\section{A No-go theorem for large $N_{c}$ QCD}

In this section, we show that the result derived in the previous section
for the Skyrme model holds for large $N_{c}$ QCD itself. The precise
statement is that if large $N_{c}$ QCD is in a phase in which the
chiral condensate $\langle\overline{q}q\rangle$ is generally non-zero
but varies from point to point, then it is not possible for all chiral
order parameters to vanish under spatial averaging. The strategy for
doing this exploits the fact that the proof in the previous section
holds for any field configurations describable as a nonlinear sigma
model. Thus our theorem will be established in general provided we
can show rigorously that in a putative chirally-restored phase with
a spatially varying but nonzero chiral condensate the expectation
values for a key set of operators are reducible to those of a nonlinear
sigma model.

To do this we focus on the scalar-isoscalar and pseudoscalar-isovector
quark bilinears $\overline{q}q$ and $\overline{q}\overrightarrow{\tau}\gamma^{5}q$.
These operators transform into one another as members of a ($\frac{1}{2}$,$\frac{1}{2}$)
representation of the ${\rm SU_{L}}(2)\times{\rm SU_{R}}(2)$ chiral
group. Suppose that the system is in some known quantum mechanical
state. Let us combine the expectation values of these operators in
this state into a single two-dimensional matrix-valued function: 
\begin{equation}
V(\mathbf{x})\equiv\frac{\left\langle \overline{q}(\mathbf{x})q(\mathbf{x})\right\rangle \, I\,+\,\left\langle \overline{q}(\mathbf{x})\overrightarrow{\tau}\gamma^{5}q(\mathbf{x})\cdot\overrightarrow{\tau}\right\rangle }{\langle\overline{q}q\rangle_{{\rm vac}}}
\end{equation}
 where $I$ is the identity matrix and the normalization factor, $\langle\overline{q}q\rangle_{{\rm vac}}$
is included so that in the vacuum state $V$ simply becomes the identity
matrix. Note that under chiral transformations on the operators making
up $V$ means it transforms in precisely the same way as $U$ does
in the Skyrme model. One can construct order parameters from $V$.
Since we are at large $N_{c}$, 
\begin{equation}
\left\langle F\left(V(\mathbf{x})\right)\right\rangle =F\left(\left\langle V(\mathbf{x})\right\rangle \right)\;,
\end{equation}
where the angle brackets indicate quantum expectation values in the
state. Thus we can treat the issue of chiral restoration in the spatially-averaged
sense as if $V$ were a classical function.

To proceed further let us introduce a chiral scalar function given
by the norm of $V$: 
\begin{equation}
v(\mathbf{x})=\sqrt{\frac{{\rm Tr}\left(V^{\dagger}(\mathbf{x})V(\mathbf{x})\right)}{2}}\;.
\end{equation}
By construction, $v(\mathbf{x})$ is real and nonnegative. For all
points in space where $v(\mathbf{x})\neq0$ one can define 
\begin{equation}
U(\mathbf{x})\equiv\frac{V(\mathbf{x})}{v(\mathbf{x})}
\end{equation}
 Note that this construction ensures that $U(\mathbf{x})\in{\rm SU}(2)$.
Thus provided that $v(\mathbf{x})$ is non-zero throughout space and
the expectation values vary smoothly through space, we have an associated
smoothly varying SU(2) matrix $U(\mathbf{x})$, precisely as in nonlinear
sigma models such as the Skyrme model. Note that the argument in subsect.
\ref{2a} does not depend on the dynamics but merely on the transformational
properties plus assumptions about smoothness. Thus we conclude that
\textit{if} the system is in a phase which is chirally restored in
the average sense while simultaneously having $v(\mathbf{x})\ne0$
everywhere, all chiral scalars must be spatially uniform. Note moreover
that $v(\mathbf{x})$ is a chiral scalar itself and therefore must
itself be a constant in any putative chirally restored phase with
$v$ everywhere non-zero.

Next, we need to eliminate the possibility in large $N_{c}$ QCD of
$v(\mathbf{x})$ vanishing at some points in space while being in
a phase which has a spatially varying chiral condensate while being
chirally restored in the spatially-averaged sense. To proceed let
us consider what happens if $V$ is distorted slightly away from its
physical value: 
\begin{equation}
\tilde{V}(\mathbf{x})=V(\mathbf{x})+\lambda\,\Delta V(\mathbf{x})
\end{equation}
 where $\lambda\,$ is a small parameter and $\Delta V$ is an arbitrary
smooth function with the property that it is non-zero in the neighbor
of any point where $V$ vanishes. We will use a tilde to denote a
quantity perturbed away from its physical value due to the use of
$\tilde{V}$ in place of $V$. Note that the change in $V$ due to
this distortion is small and smooth. So the changes in spatially-averaged
chiral order parameters obtained from $\tilde{V}$ are expandable
as a Taylor series in $\lambda$. Thus \textit{if} the system was
in a putative phase, which was chirally restored in the average sense,
then spatially-averaged chiral order parameters computed with $\tilde{V}$
would be of order $\lambda\,$; $\tilde{V}$ would describe a nearly
chirally restored regime.

To proceed, we exploit the fact that $\tilde{V}$ is constructed to
be nonzero everywhere, implying that $\tilde{U}$ is well-defined
everywhere. We will describe any putative nearly chirally restored
regime associated with $\tilde{V}$ using $\tilde{U}$. Since the
supposed regime is \textit{nearly} chirally restored, the argument
in subsect. \ref{2a} goes through up to corrections associated with
the fact that it is only nearly restored. Thus we see that in such
a regime all chiral singlets must be nearly constant in space: $\tilde{s}(\mathbf{x})=\tilde{s}_{0}+\lambda\,\delta\tilde{s}(\mathbf{x})$,
where $\tilde{s}_{0}$ is a constant and $\lambda\,\delta\tilde{s}(\mathbf{x})$
describes the fluctutations away from it. Since $\tilde{v}(\mathbf{x})$
is a chiral scalar we conclude that $\tilde{v}(\mathbf{x})=\tilde{v}_{0}+\lambda\,\delta\tilde{v}(\mathbf{x})$.
On the other hand, by construction at any point in space, $\tilde{v}(\mathbf{x})$
can be expressed as a Taylor expansion in $\lambda\,$: $\tilde{v}(\mathbf{x})=v(\mathbf{x})+\lambda\,\Delta v(\mathbf{x})$.
Equating these two forms yields 
\begin{equation}
v(\mathbf{x})=\tilde{v}_{0}+\lambda\,\left(\delta\tilde{v}(\mathbf{x})-\Delta v(\mathbf{x})\right)\;.
\end{equation}
 This implies that as $\lambda\,\rightarrow0$, $v(\mathbf{x})$ approaches
a constant value. However, a constant value of $v(\mathbf{x})$ is
inconsistent with $v$ going to zero at some points while generally
being non-zero as a result of the non-zero chiral condensate. Thus
\textit{if} a phase which is chirally restored in the spatially averaged
sense exists and has a spatially varying chiral condensate, it must
have a constant non-zero $v(\mathbf{x})$. Therefore, it is effectively
reduced to a nonlinear sigma model and the proof in the preceding
section applies. This completes the demonstration.

\section{Skyrmions on a Hypersphere}

We have shown that chiral symmetry restoration in the spatially-averaged
sense is not possible for either the Skyrme model or for large $N_{c}$
QCD. However, two important issues arise regarding the Skyrme model.
As was noted in the introduction observed chiral restoration in the
spatially-averaged sense \textit{is observed} for the Skyrme model
on the hypersphere \cite{key-13,key-2,key-6,key-9,key-11}.

The first issue is quite straightforward. One might worry that since
chiral symmetry does get restored in the average sense for Skyrmions
on the hypersphere this might indicate there is a flaw in the proof
that Skyrmions cannot have chiral restoration at finite density. Of
course, it need not indicate a flaw since the geometry for which situation
for which the no-go theorem was derived---extended nuclear matter
in flat space---is different from the curved space of the hypersphere.
Nevertheless, it is important to pin down how the proof can be valid
for three-dimensional flat space but fail for the hypersphere.

The second issue concerns intuition. As noted in the introduction,
the approximation of a Skyrmion crystal by a single Skyrmion on a
hypersphere is \textit{ad hoc}. The notion, though, was that the main
qualitative effect of putting a Skyrmion into a crystal is to restrict
the space over which it can spread. This effect is certainly also
present with a single Skyrmion on a hypersphere. To the extent that
qualitative questions such as whether or not chiral restoration (in
a spatially-averaged sense) occurs is a generic feature which depends
on whether a Skyrmion is confined to a sufficiently small volume,
one might think that a Skyrmion on a hypersphere would be a good way
to discover this. However, this is clearly wrong. Chiral restoration
in the average sense occurs on the hypersphere but not for crystals
in flat space. It is important to understand why the intuition gained
from the hypersphere fails.

Before addressing these issues, let us briefly review the geometry
of the hypersphere. This can be characterized as a three-dimensional
curved surface in a four-dimensional Euclidean space with fixed radius:
\begin{equation}
x^{2}+y^{2}+z^{2}+w^{2}=L^{2}\label{hs}
\end{equation}
 where $L$ is the radius of the hypersphere. It is useful to parameterize
this in terms of three angular variables, $0\leq\mu,\theta\leq\mbox{\ensuremath{\pi}}$
and $0\leq\phi\leq2\mbox{\ensuremath{\pi}}$ with 
\begin{equation}
\begin{split} & x=L\sin\mu\sin\theta\cos\phi\\
 & y=L\sin\mu\sin\theta\sin\phi\\
 & z=L\sin\mu\cos\theta\\
 & w=L\cos\mu\end{split}
\end{equation}

We denote the metric tensor for this geometry as $g^{{\rm hs}}$.
With parameterization used here it is diagonal and the diagonal matrix
elements are: \begin{equation}
\begin{split}g_{\mu\mu} & =L^{2}\\
g_{\theta\theta} & =L^{2}\sin^{2}\mu\\
g_{\phi\phi} & =L^{2}\sin^{2}\mu\sin^{2}\theta\,.\end{split}
\end{equation}
 The volume element is given by \begin{equation}
dV=\sqrt{\det g\,}d\mu\, d\theta\, d\phi=L^{3}\sin^{2}\mu\, d\mu\sin\theta\, d\theta\, d\phi\end{equation}

Consider the configuration 
\begin{equation}
\begin{split} & U_{0}(\mu,\theta,\phi)=\\
 & \left(\begin{array}{ccc}
\cos(\mu)+\sin(\mu)\cos(\theta) & \sin(\mu)\sin(\theta)\exp(-i\phi)\\
\sin(\mu)\sin(\theta)\exp(i\phi) & \cos(\mu)-\sin(\mu)\cos(\theta)\end{array}\right)\;.\end{split}
\label{U0}
\end{equation}
 It is straightforward to verify that this configuration has winding
number unity and that any chiral order parameter constructed from
$U_{0}$ vanishes when integrated uniformly over the hypersphere.
Thus the configuration $U_{0}$ corresponds to chiral restoration
in the spatially averaged sense on the hypersphere. The explicit example
of $U_{0}$ shows that the hyperspherical geometry does allow configurations
with chiral restoration in the spatially-averaged sense. The theorem
proved in Sect. \ref{2} showed such a configuration---at least for
the case of flat space. This means either the proof of the theorem
is wrong or some aspect of the proof holds for flat space but fails
for the case of the hypersphere.

Fortunately, it is easy to see how the proof of the no-go theorem
can hold for flat space and not for the hypersphere. At several critical
point points in the derivation the fact that the space is flat plays
a critical role. Consider as an example the relation $\overleftrightarrow{R}(\mathbf{x}+\mathbf{d})=\overleftrightarrow{R}(\mathbf{x})\overleftrightarrow{R}(\mathbf{d}+\mathbf{x}_{0})$
from Eq.~(\ref{RR}); this relation is central in the analysis leading
to Eq.~(\ref{com}) which is at the core of the theorem. However,
as written this result is not meaningful on the hypersphere: there
is no notion of linearly adding the vectors associated with two points
in the space to obtain the vector associated with a third point in
the space. Thus the notion of the point $\mathbf{x}+\mathbf{d}$ is
simply ill-posed on the hypersphere and the proof developed in flat
space does not go through.

Let us now turn to the issue of the failure of the intuition that
a single Skyrmion in the hypersphere should be qualitatively similar
to the Skyrme crystal as the principal effects of both should be to
confine a Skyrmion to a limited spatial region. The intuition can
break down for one of two reasons: either limiting the space in which
a Skyrmion extends by a single Skyrmion in a compact geometry is \textit{not}
qualitatively similar to limiting it by placing in a crystal in flat
space, or because the hypersphere is an atypical compact geometry.
It is instructive to understand which is the cause. There are good
reasons to suspect that it is due the atypical properties of the hypersphere.
The hypersphere has a much higher symmetry than typical geometries
one can consider. It is plausible that these symmetries rather than
generic properties are responsible for the chiral restoration (in
the spatially-averaged sense) seen for Skyrmions on the hypersphere.
This becomes particularly plausible when one considers the internal
space associated with the field $U$. It can be parameterized as \begin{equation}
U=sI+i\vec{p}\cdot\vec{\tau}\label{uparm}\end{equation}
 where $s$ and $\vec{p}$ can be extracted from $U$ using the relations
\begin{equation}
s=\frac{{\rm Tr}\left(U\right)}{2}\;\;,\; p_{j}=-\frac{-i{\rm Tr}\left(\tau_{j}U\right)}{2}\;.\label{sp}
\end{equation}
 Note that by construction $s$ and $\vec{p}$ satisfy the constraint
\begin{equation}
s^{2}+p_{1}^{2}+p_{2}^{2}+p_{3}^{2}=1\;.\label{constr}
\end{equation}
 This constraint in the internal geometry in Eq.~(\ref{constr})
is of the same form as the constraint in Eq.~(\ref{hs}) describing
the hypersphere. Indeed it is easy to see that the configuration in
Eq.~(\ref{U0}) which restores chiral symmetry on the hypersphere
(in the spatially averaged sense) has the property that it maps points
on the hypersphere to the analogous points in the SU(2) matrix $U$:
\begin{equation}
s=\frac{w}{L}\,,\;\; p_{1}=\frac{x}{L}\,,\;\; p_{2}=\frac{y}{L}\,,\;\; p_{3}=\frac{z}{L}\,.
\end{equation}
 Thus averaging over the hypersphere uniformly automatically leads
to averaging uniformly over the internal space forcing all chiral
order parameters constructed from $U$ to vanish.

One way to test whether the observed chiral restoration seen for a
Skyrmion on a sufficiently small hypersphere is due to the special
geometrical properties of the hypersphere is to ask what happens to
spatially-averaged chiral order parameters in this regime if the geometry
is distorted slightly away from a hypersphere. If chiral symmetry
is generically restored for single Skyrmions in small compact spaces,
these spatially averaged order parameters should remain zero. If instead
chiral restoration is a result of special features associated with
the hyperspherical geometry one would expect these to become non-zero.

To make this analysis concrete, we need to impose dynamics associated
with a particular variant of the Skyrme model. Here, we will use the
simplest one which contains field $U$ and up to four-derivative terms:
\begin{equation}
\mathcal{L}=\frac{f_{\pi}^{2}}{4}Tr[\partial_{\mu}U^{\dagger}\partial^{\mu}U]+\frac{\epsilon^{2}}{4}Tr[U^{\dagger}\partial_{\mu}U,\, U^{\dagger}\partial_{\nu}U]^{2}\,,\label{lag}
\end{equation}
where $f_{\pi}\approx93MeV$ is the pion decay constant and $\epsilon$
is a dimensionless parameter necessary to stabilize the soliton. In
the remainder of this paper, we will use dimensionless units by setting
$f_{\pi}$ and $\epsilon$ to unity. One can use dimensional analysis
to reinsert factors of $f_{\pi}$ and $\epsilon$ at the end of the
problem, if desired. To do so, simply multiply lengths by $2\sqrt{2}\epsilon/f_{\pi}$
and energies by $\sqrt{2}\epsilon f_{\pi}$ \cite{key-2}. For the
purposes of determining whether or not chiral symmetry is restored,
these rescaling factors are irrelevant.

For the case of an unperturbed hyperspherical geometry, the minimum
energy winding number unity configuration is a hedgehog localized
in one region of the hypersphere, providing the radius $L$ is smaller
than the critcal value of $\sqrt{2}$ \cite{key-9}. For $L<\sqrt{2}$,
the minimum energy configuration is given by $U_{0}$ in Eq.~(\ref{U0})
or configurations obtained from it by a global SU(2) rotation in internal
space. Our strategy is to start with the lowest energy configuration
for a value of $L$ well into the restored phase (for concreteness
we take $L=1$). We then consider a small perturbation on the geometry,
compute the shift in the configuration due to this to lowest order
in the perturbation, and use this perturbed configuration to compute
the spatially-averaged chiral order parameters at first order in the
shift. If chiral restoration in the spatially-averaged sense were
a generic feature seen when a Skyrmion is confined by a compact geometry
of sufficiently small size one expects spatially averaged order parameters
to remain zero.

The most general perturbation of the hypersphere can be formulated
as a shift in the contravariant metric perturbation away from that
of the hypersphere parameterized in terms of an overall expansion
parameter $\lambda\,$: 
\begin{equation}
g_{pert}^{{\rm {\it ij}}}=g_{hs}^{{\rm {\it ij}}}+\frac{\lambda}{L^{2}}\,\gamma^{{\it ij}}\,,\label{pmatrix}
\end{equation}
where $\gamma^{ij}$ is dimensionless. We will consider a very simple
class of perturbations to the geometry: those in which all matrix
elements of $\gamma$ vanish except the $(i=1,\, j=1)$ component.
Further, we will take $\gamma^{11}$ to be a sinusoidal function of
$\mu$ onl,y which ensures that the perturbation does not induce discontinuities
in derivatives: 
\begin{equation}
\gamma^{ij}=\delta^{i1}\,\delta^{j1}\sin(p\mu)\,,
\end{equation}
where $p$ is a non-zero integer. Of course, this is a very restricted
class of perturbation. If spatially-averaged chiral order parameters
remain zero with it, it would tell us very little since this might
be a result of the highly constrained and symmetric form of the metric
perturbation. However, if they do not remain zero, then it is sufficient
to show that the hyperspherical geometry does not behave generically.

We now solve the Skyrme model for the lowest energy unit winding number
solution with this geometry working to first order in $\lambda$.
To do so we consider configurations of the most symmetric sort consistent
with a winding number of unity. These are hedgehog configurations
of the form 
\begin{equation}
\begin{split} & U(\mu,\theta,\phi)=\exp\left(i\vec{\tau}\cdot\hat{r}f(\mu)\right)\;\;{\rm where}\;\hat{r}=\left(\begin{array}{c}
\sin(\theta)\,\cos(\phi)\\
\sin(\theta)\,\sin(\phi)\\
\cos(\theta)\end{array}\right)\\
 & {\rm with}\;\; f(0)=0\;\;{\rm and}\; f(\pi)=\pi\;.\end{split}
\label{ansatz}
\end{equation}
We note that for the unperturbed geometry, $U_{0}$, the configuration
in Eq.~(\ref{U0}) associated with chiral restoration is of this
form with $f(\mu)=\mu$. For the perturbed case, we exploit the fact
that the most symmetric form remains consistent with the topological
constraints. It is guaranteed that the configuration of this form,
which minimizes the energy, will also be a solution of the full Euler-Lagrange
equations. As a general rule, a solution of the Euler-Lagrange equations
obtained from a highly symmetric ansatz need not correspond to a global
minimum of the energy. However in this case it should, as in the unperturbed
case where the hedgehog is known to be a global minimum for that problem
and this problem differs from that only to first order in $\lambda$.
Next we take the ansatz in Eq.~(\ref{ansatz}) using 
\begin{equation}
f(\mu)=\mu+\lambda\delta f(\mu)\;\;\;{\rm with}\;\delta f(0)=\delta f(\pi)=0\;.
\end{equation}
Using this form, we compute the energy to second order in $\lambda$
(noting that first-order perturbations in the field correspond to
second-order perturbations in the energy) and vary this form to get
differential equations for $\delta f$ which are then solved numerically.
Using these numerical solutions the spatial averages of various chiral
order parameters are computed.

A useful set of chiral order parameters to consider are local functions
constructed from $U$ or, equivalently, from $s$ and $\vec{p}$ from
Eq.~(\ref{sp}). We will focus here on three representative samples
from this list: 
\begin{equation}
s\,,\;\;\frac{3s^{2}}{4}-\frac{p_{1}^{2}+p_{2}^{2}+p_{3}^{2}}{4}\;\;{\rm and}\; s^{3}\,.
\end{equation}
It is clear that $s$ and $s^{3}$ are chiral order parameters since
$s\rightarrow-s$ is a chiral transformation. Similarly, it should
be clear that $\frac{3s^{2}}{4}-\frac{p_{1}^{2}+p_{2}^{2}+p_{3}^{2}}{4}$
is a chiral order parameter since in a chirally restored phase one
necessarily has $\langle s^{2}\rangle=\langle p_{1}^{2}\rangle=\langle p_{2}^{2}\rangle=\langle p_{3}^{2}\rangle$
and thus the quantity vanishes. In Table \ref{T1}, we show the spatial
averages of these order parameters computed to leading order in $\lambda$.

\begin{table}
\caption{Spatially-averaged chiral parameters at order $\lambda$. These are
computed using geometries of the perturbed hypersphere given in Eq.~(\ref{pmatrix}).
The configurations used to compute these are minimum energy configurations
of the hedgehog form based on the Lagrangian in Eq.~(\ref{lag}). }

\label{T1} \begin{tabular}{|c|c|c|c|}
\hline 
metric perturbation  & $\;\;\; s\;\;\;$  & $\frac{3s^{2}}{4}-\frac{p_{1}^{2}+p_{2}^{2}+p_{3}^{2}}{4}$  & $s^{3}$\tabularnewline
\hline
\hline 
$\gamma^{ij}=\delta^{i1}\,\delta^{j1}\sin(\mu)$  & 0  & -0.0954931 $\lambda$  & 0\tabularnewline
\hline 
$\gamma^{ij}=\delta^{i1}\,\delta^{j1}\sin(2\mu)$  & 0  & 0  & -0.0314619 $\lambda$ \tabularnewline
\hline 
$\gamma^{ij}=\delta^{i1}\,\delta^{j1}\sin(3\mu)$  & 0  & -0.0197049 $\lambda$  & 0 \tabularnewline
\hline 
$\gamma^{ij}=\delta^{i1}\,\delta^{j1}\sin(4\mu)$  & 0  & 0  & -0.00699138 $\lambda$ \tabularnewline
\hline 
$\gamma^{ij}=\delta^{i1}\,\delta^{j1}\sin(5\mu)$  & 0  & 0.0156628 $\lambda$  & 0 \tabularnewline
\hline
\end{tabular}
\end{table}

It is noteworthy that not all entries in this table are zero. Thus
making small perturbations away from the hyperspherical geometry yields
non-zero order parameters. From this we see that chiral restoration
seen for Skyrmions on the hypersphere is not generic but rather is
a consequence of the special geometric properties of the hypersphere.
Of course, this geometry has no physical significance and was used
for ease of computation. While it was hoped that a calculation in
this simple situation would give useful physical intuition about the
more complicated situation of Skyrmion crystals, it appears that the
opposite is true with regard to chiral symmetry. Indeed, the special
properties of the geometry which makes analytic calculations of the
small radius (high density) phase simple also make the intuition totally
unreliable even for qualitative issues associated with chiral symmetry
breaking and its possible restoration in the average sense.

\section{Discussion}

The question of whether dense, cold nuclear matter is chirally restored
above some critical baryon chemical potential remains a problem of
central importance to nuclear physics. It remains unanswered. The
answer is also potentially unknown in the simpler case of large $N_{c}$
QCD.

This paper dealt with a complication at large $N_{c}$: the fact that
nuclear matter is likely to crystallize and thus the breaking of chiral
symmetry and of translational symmetry may get entangled. One could
imagine a situation in which the breaking of translational symmetry
can introduce non-zero but spatially-averaged chiral order parameters
which integrate to zero over space. To deal with this complication,
it was suggested here that the natural focus should be on spatially-averaged
chiral order parameters which by construction are insensitive to details
of how translational symmetry is broken. We demonstrated here that
while it is possible for some spatially-averaged chiral order parameters
to vanish, it is not possible for all of them to if the chiral condensate
is generally non-zero but spatially varying. Thus chiral symmetry
restoration in a spatially averaged sense is not possible at large
$N_{c}$ unless the chiral condensate is zero everywhere. 

This result is of some significance in connection to the question
of whether chiral restoration occurs at high baryon density in large
$N_{c}$ QCD. It has been argued on the basis of Skyrme models that
chiral restoration does occur at sufficiently high density. The fact
that at high densities the lowest energy configurations in Skyrme
crystals have a vanishing spatially-averaged chiral condensate has
be taken as support for chiral restoration \cite{key-3}. As shown
here, however, there must be other chiral order parameters which are
non-zero upon spatial averaging. It has similarly been argued that
the vanishing of chiral order parameters upon spatial averaging for
a Skyrmion on a sufficiently small hypersphere has also been taken
as evidence for chiral restoration at high density \cite{key-3}.
However, as shown here this was an artifact of the hyperspherical
geometry, and is thus unconnected to the question of what happens
in crystals in flat space.

\mbox{}

T.D.C. and P.A. acknowledge the support of the U. S. Department of
Energy through grant number DEFG02-93ER-40762. M.S. is grateful for
the support from the University of Maryland Office of Undergraduate
Studies and R.A. acknowledges the support of the Theoretical Quarks,
Hadrons and Nuclei (TQHN) group at the University of Maryland.

\end{document}